\documentstyle[aps,preprint,epsf]{revtex}
\newcommand{\be}{\begin{equation}}
\newcommand{\ee}{\end{equation}}
\newcommand{\bear}{\begin{eqnarray}}
\newcommand{\eear}{\end{eqnarray}}

\def\fun#1#2{\lower3.6pt\vbox{\baselineskip0pt\lineskip.9pt
 \ialign{$\mathsurround=0pt#1\hfil##\hfil$\crcr#2\crcr\sim\crcr}}}

\begin{document}
\draft
%\wideabs{
\title{STRANGENESS PHOTOPRODUCTION FROM THE DEUTERON AND
HYPERON--NUCLEON INTERACTION}
 \author{B. Kerbikov}
\address{ Institute of Theoretical and Experimental
Physics, 117218, Moscow, Russia}
% \date{}

\maketitle

\begin{abstract}

Pronounced effects due to final state hyperon--nucleon interaction
are predicted in strangeness photoproduction reaction on the
deuteron. Use is made of the covariant reaction formalism and the
$P$-- matrix approach to the hyperon--nucleon interaction.


   \end{abstract}

%%%%%%%%%%%%%%%%%%%%%%%%%%%%%%%%%%%%%%%%%%%%%%%%%%%%%%%%%%%%%%%%%%%%%%%%

The  first measurement of kaon photoproduction on  the deuteron are
anticipated this year [1]. These data may open a new window on the
$\Lambda N$ and $\Sigma N$ forces since the final state $YN$
interaction (FSI) plays an important role in the $\gamma d\to K^+YN$
reaction. This problem has been addressed by several authors starting
from the pioneering paper by  Renard and Renard [2-3]. Two points make
the present work different from the previous studies: {\em (i)} the
use of the covariant formalism both for the reaction mechanism and
the deuteron wave function, and {\em (ii)} the $P$--matrix approach
to the FSI which takes into account the subnuclear degrees of freedom
and disentangle the dynamical singularities from kinematical
threshold effects [4]. Our main result is a prediction of the
spectacular effects in the  reaction cross section due to the $YN$
FSI.

The reaction $\gamma d \to K^{+} Y n$, $Y = \Lambda, \Sigma^{0}$
is a $2 \to 3$ process. The corresponding double
differential cross--section reads
\begin{equation}
	d^2 \sigma \equiv \frac{d^2 \sigma}{d\vert {\bf p}_{K} \vert d\Omega_{K}}
	= \frac{1}{2^{11}\pi^{5}} \frac{{\bf p}_{K}^{2}}{k M_{d}E_{K}}
	  \frac{\lambda^{1/2}(s_{2}, m_{Y}^{2}, m_{n}^{2})}{s_{2}}
	  \int d\Omega^{*}_{Yn} \vert T\vert ^{2} \; .
	\label{eq:dda-sigma}  % (2.3)
\end{equation}

Here $k$, ${\bf p}_{K}^{2}$, $E_{K}$ and $\Omega_{K}$ correspond to
the deuteron rest system  with $z$-axis defined by the incident
photon beam direction ${\bf k}$. The solid angle $\Omega^{*}_{Yn}$ is
defined in the $Yn$ center-of-momentum system. The quantity
$\lambda(x, y, z)$ is the standard kinematical function
$\lambda(x, y, z) = x^{2} - 2(y+z)x + (y-z)^{2}$.

We shall use the covariant relativistic approach to calculate the
amplitude $T$ of the process $ \gamma d\to K^+YN$.
 The amplitude will be approximated by the  two  leading diagrams,
 namely the tree (pole, or plane waves) graph and the triangle graph
 with FSI.  It will be demonstrated
 that within  the covariant approach one easily retrieves the usual
 nonrelativistic impulse approximation and the
Migdal-Watson approach to FSI.
We start with the tree
diagram.
To calculate it two blocks have to be
specified:  {\em (i)} the elementary photoproduction amplitude
$M^{\gamma K}$ on the
proton, and {\em (ii)} the deuteron vertex $\Gamma_d$.
The elementary amplitude used in the present calculation was derived
from the tree level effective Lagrangian [5]. Taken into account were
resonances with the spin $\leq 5/2$ in the $s$--channel the spin--1/2
resonances in the $u$--channel, and $K^*(892)$ and $K1(1270)$
resonances in the $t$-- channel.  This amplitude has the following
decomposition over invariant terms [6]
 \begin{equation}
  M^{\gamma K}
= \overline{u}_{Y} \sum_{j=1}^{6} {\cal A}_j {\cal M}_{j}(s', t', u')
u_{p} \; , \label{eq:mgk}
 % (3.1)
  \end{equation}
  where $s'  =
(k+p_{p})^{2} $, $t'  = (k-p_{K})^{2}$, $u' = (k-p_{Y})^{2}$.

 The decomposition of the deuteron  vertex function $\Gamma_{d}$ in
independent Lorentz structures has the form [7]

\begin{eqnarray}
\Gamma_{d}       & = & \sqrt{m_{N}}
\left[ (p_{p} + p_{n})^{2} -M_{d}^{2} \right]
     \left[\varphi_{1}(t_{2}) \frac{(p_{p}-p_{n})_{\mu}}{2m_{N}^{2}}
     + \varphi_{2}(t_{2}) \frac{1}{m_{N}} \gamma_{\mu} \right] {\cal
            E}^\mu(p_{d}, \lambda).
\end{eqnarray}

Here $t_2=(p_d-p_n)^2,$
 ${\cal E}^\mu(p_{d}, \lambda)$ is the polarization
4-vector of the deuteron with momentum $p_{d}$ and polarization
$\lambda$.

Now we can  write  the following expression for the tree diagram

\begin{equation}
	T^{(t)} = \overline{u}_{Y}
	\left\{ \left (\sum_{j=1}^{6} {\cal A}_{j} {\cal M}_{j} (s', t', u')
	\right )
	S(p_{p}) \Gamma_{d} \right\} u_{n}^c \; ,
	\label{eq:treeone}  % (3.10)
\end{equation}

\noindent where $u_{n}^c$ is a charge conjugated neutron spinor.

Covariant equations (1) and (4) can be easily reduced to the standard
impulse approximation. Neglecting the spin summation in the matrix
element (4) (factorization conjecture) and introducing the deuteron
wave function as a product [8] $\Psi_d=[2(2\pi)^3 M_d]^{1/2} S(P_p)
\Gamma_d$, one retrieves the nonrelativistic impulse approximation
\begin{equation}
       \frac{ d^{2}\sigma}{d|{\bf p}_K|d\Omega_K} =
        \frac{{\bf p}_{K}^{2} \vert {\bf p}_{Y}^{*}
        \vert}{64\pi^{2}k E_{K} \sqrt{s_{2}}} \int d\Omega^{*}_{Yn}
        \; \vert M^{\gamma K} \vert^{2} \vert \psi_{d} \vert^{2}
        \; , \label{eq:dtwo}   % (3.17)
         \end{equation}

\noindent where ${\bf p}_{Y}^{*}$ corresponds to the $YN$
center--of--momentum  system.
The main physical difference between the covariant deuteron vertex
used in the present calculation and the nonrelativistic wave function
entering into (5) is that the former contains singlet and triplet
$p$--wave components absent in the later [8].

Next consider the loop (triangle) diagram  with $YN$~~~
FSI. The corresponding amplitude is given by

\begin{equation}
	T^{(l)} = \int \frac{d^{4} p_{n}}{(2\pi)^{4}} \;
	\overline{u}_{Y}(p'_{Y})
	\left\{ \left( \sum_{j=1}^{6} {\cal A}_{j} {\cal M}_{j} \right)
	 S(p_{p})  \Gamma_{d} C S(p_{n}) T_{Yn} S(p_{Y}) \right\}
	 \overline{u}(p'_{n}) \; .
	\label{eq:tl1}  % (4.1)
\end{equation}

Here $C = \gamma_{2}\gamma_{0}$ is the charge-conjugation matrix,
$T_{Yn}$ is a four-fermion hyperon-nucleon vertex,
this vertex
being ``dressed'' by corresponding spinors constitutes the
hyperon-nucleon amplitude $F_{Yn}$.

The comprehensive treatment of the loop diagram will be presented in
the forthcoming detailed publication while here we resort to a simple
approximation with the aim to exposure the effects of the FSI.
Namely, only positive frequency components are kept in all three
propagators $S(p_j)$, $j=p,n,Y$, then the integration over the time
component $dp^0_n$ is performed and the deuteron wave function is
introduced in the same way as it was done in arriving to equation
(5).
Thus we get  the following expression for $T^{(l)}$:

\begin{equation}
T^{(l)} = - \sqrt{(2\pi)^{3} 2 M_{d}}
\int \frac{d{\bf p}^{*}}{(2\pi)^{3}}
\frac{M^{\gamma K} \psi_{d} F_{Yn}(E^{*}_{Yn})}%
{{\bf p}^{*2} - {\bf p'}^{*2} -i0} \; ,
	\label{eq:tl2}  % (4.2)
\end{equation}

\noindent where ${\bf p}^{*2}$ and ${\bf p'}^{*2}$ are the $Yn$
momenta in the center--of--mass $Yn$ system before and after the
rescattering, $F_{Yn}(E^{*}_{Yn})$ is the half-off-shell $Yn$
scattering amplitude at the energy $E^{*}_{Yn} = {\bf
p'}_{Yn}^{*2}/m_{Yn}$.
The use of the nonrelativistic propagator in (7) is legitimate since
FSI is essential at
low $YN$ relative momenta. In the kinematical region where FSI is
 important the amplitude $M^{\gamma K}$ and the  deuteron wave
 function $\psi_d$ are smoother functions of ${\bf p}^*$ compared to
the scattering amplitude $F_{Yn}$. Therefore one can set ${\bf
p}^*={\bf p}^{'*}$ in their arguments and take them off the integral.
Next recall that as it was shown above the tree (plane waves)
amplitude $T^{(t)}$ allows the representation

\begin{equation}
T^{(t)} \simeq \sqrt{(2\pi)^{3} 2 M_{d}}\;  M^{\gamma K} \psi_{d} \; .
	\label{eq:tt1}  %  (4.4)
\end{equation}

Therefore for the sum of the two diagrams we can write

\begin{equation}
T^{(t)} + 	T^{(l)} = T^{(t)} \int \frac{d{\bf p}^{*}}{(2\pi)^{3}}
             \Psi_{{\bf p}^{*'}}^{(-)*} ({\bf p}^{*})
             \equiv
             T^{(t)} /D,
\end{equation}

\noindent where

\begin{equation}
	\Psi_{{\bf p}^{*'}}^{(-)} ({\bf p}^{*}) =
	(2\pi)^{3} \delta ({\bf p}^{*}-{\bf p}^{*'}) -
        \frac{F_{Yn}(E^{*}_{Yn})}{{\bf p}^{*2} - {\bf p'}^{*2} +i0}
        \; .  \label{eq:bigpsi}
        % (4.6)
         \end{equation}
         and $1/D$
denotes the enhancement factor which will be calculated in the
$P$--matrix approach. The $P$--matrix description of the $YN$
interaction including threshold phenomena and the resonance at 2.13
GeV close to the $\Sigma N$ threshold
was presented in [9]. According to [9] the 2.13 GeV
structure is not a genuine six--quark state but the $P$--matrix
partner of the deuteron
(see also [10]). Using the connection between
$S$-- and $P$-- matrices
[11], one arrives at the following expression for
the enhancement factor
\be
D^{-1}=\frac{e^{-ip^{'*}b}}{2ip^{'*}b}\left\{
\frac{e^{-ip^{'*}b}R(-ip^{'*})-
e^{ip^{'*}b}R(+ip^{'*})}{R(+ip^{'*})}\right\},
\ee
where
\be
R(ip^{'*})=(P^0_1-ip^{'*})(E-E_n+
\frac{\Lambda^2_1}{P_1^0-ip^{'*}}+
\frac{\Lambda^2_2}{P_2^0-ip^{'*}_2}),
\ee
and where the elements of the $P$--matrix are given by [11]
\be
P_{ij}=P^0_i\delta_{ij}+
\frac{\lambda_i\lambda_j}{E-E_n},
\ee
and indices 1 and 2 correspond to $\Lambda n$ and $\Sigma^0n$
channels, $p_2^{'*} $ is the momentum in the $\Sigma^0n$ channel.
The numerical values of the $P$--matrix parameters entering into
(11)-(13) may be found in [9].

Finally we present the results of the calculations obtained using the
equations (1),(4),(9) and (11). Use has been made of the elementary
photoproduction amplitude from [5] and the deuteron vertex function
taken from the relativistic Gross model [8]. Calculations of the
plane--waves diagram (4) with this input were performed in [12]. In
Fig.1 the double differential cross section (1) is shown as a
function of the photon energy in the $\Lambda n$ invariant mass
region close to the $\Lambda n$ threshold ($2.05 GeV \leq
\sqrt{s_{\Lambda n}}\leq 2.09 GeV$). The pronounced peak typical for
the FSI is seen at the $\Lambda n $ threshold.

\begin{figure}
\begin{center}
\epsfysize=10cm \epsfbox{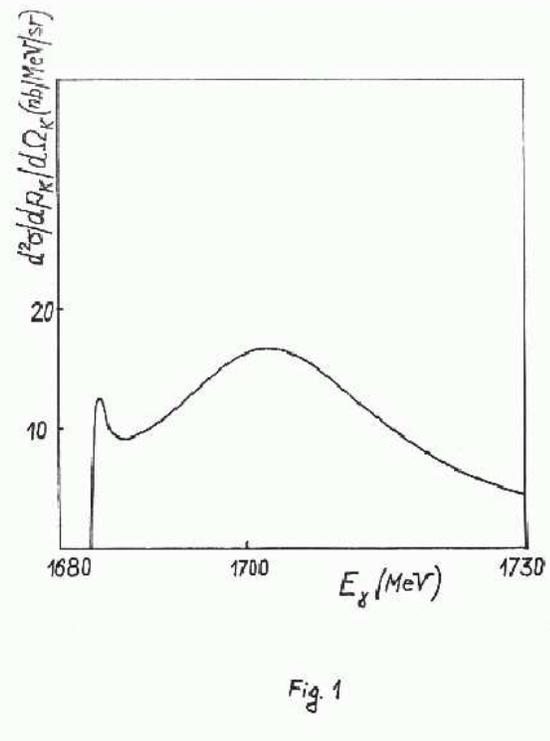}

\caption{ The double diffential cross section as a function of
         the photon energy for $P_K=1.4 GeV$ and $\theta_{\gamma K}=1^0$.}
\end{center}
\end{figure}

In Fig. 2 the same cross section is plotted in a different
kinematical conditions covering a wider region of the $NY$ invariant
mass ($2.05 GeV\leq \sqrt{s_{\Lambda n}}\leq 2.17 GeV$) including the
$\Sigma N$ threshold. Apart from the structure  at the $\Lambda n$
threshold a spectacular peak due to the $2.13 GeV$ resonance lying in
the immediate vicinity of the $\Sigma^0n$ threshold is seen. Also
shown are the results obtained with the Verma--Sural potential of the
$\Lambda N$ interaction [13]. The difference is quite distinct.

\begin{figure}
\begin{center}
\epsfysize=10cm \epsfbox{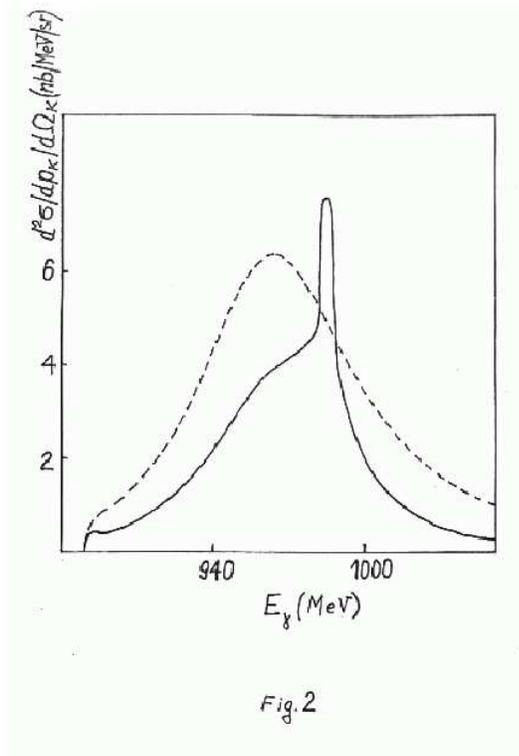}
\caption{ The double differential cross section as a function
           of photon energy for
	$P_K=0.426 GeV$ and $\theta_{\gamma
         K}=15^0$. Full -- line: $P$-- matrix result, dashed line:
	Verma--Sural potential.}
\end{center}
\end{figure}

The main conclusion is that FSI effects in the $\gamma d \to
K^+\Lambda  n$ reaction are measurable and distinctly reflect the
underlying $NY$ interaction dynamics.
Therefore detailed calculations along the lines outlined in this
paper are highly desirable. The authors would like to thank
V.A.Karmanov for fruitful discussions and suggestions. Valuable
remarks by T.Mizutani and A.E.Kudryavtsev are gratefully
acknowledged.  Special thanks are to C.Fayard,  G.H.Lamot, F.Rouvier
and B.Saghai for stimulating contacts and hospitality at all stages
of the present work. Financial support from the University Claude
Bernard, DAPNIA (Saclay) and RFFI grant 970216406 is gratefully
acknowledged.

\end{document}